\documentclass[manuscript]{aastex}

\shorttitle{Stability of sulphur dimers (S$_2$) in cometary ices}
\shortauthors{Mousis et al.}

\usepackage{etex}
\usepackage{m-pictex, m-ch-en}
\usepackage {amsmath}
\usepackage[squaren, Gray, cdot]{SIunits}
\usepackage{color}

\begin{document}


\title{Stability of sulphur dimers (S$_2$) in cometary ices}


\author{O. Mousis\altaffilmark{1}, O. Ozgurel$^{2}$, J. I. Lunine\altaffilmark{3}, A. Luspay-Kuti\altaffilmark{4}, T. Ronnet\altaffilmark{1}, F. Pauzat\altaffilmark{2}, A. Markovits\altaffilmark{2}, and Y. Ellinger\altaffilmark{2}}


\altaffiltext{1}{Aix Marseille Universit{\'e}, CNRS, LAM (Laboratoire d'Astrophysique de Marseille) UMR 7326, 13388, Marseille, France {\tt olivier.mousis@lam.fr}}
\altaffiltext{2}{Laboratoire de Chimie Th\'eorique, Sorbonne Universit\'es, UPMC Univ. Paris 06, CNRS UMR 7616, F-75252 Paris CEDEX 05, France}
\altaffiltext{3}{Department of Astronomy and Carl Sagan Institute, Space Sciences Building Cornell University,  Ithaca, NY 14853, USA}
\altaffiltext{4}{Department of Space Research, Southwest Research Institute, 6220 Culebra Rd., San Antonio, TX 78228, USA}

\begin{abstract}

S$_2$ has been observed for decades in comets, including comet 67P/Churyumov-Gerasimenko. Despite the fact that {this molecule} appears ubiquitous in these bodies, the nature of its source remains unknown. In this study, we assume that S$_2$ is formed by {irradiation (photolysis and/or radiolysis) of S-bearing molecules} embedded in {the icy grain precursors of comets}, and that {the cosmic ray flux} simultaneously creates voids in ices within which the produced molecules can accumulate. We investigate the stability of S$_2$ molecules in such cavities, assuming that the surrounding ice is made of H$_2$S or H$_2$O. We show that the stabilization energy of S$_2$ molecules in such voids is close to that of the H$_2$O ice binding energy, implying that they can only leave the icy matrix when this latter sublimates. Because S$_2$ has a short lifetime in the vapor phase, we derive that its formation in grains via {irradiation} must occur only in low density environments {such as the ISM or the upper layers of the protosolar nebula, where the local temperature is extremely low. In the first case, comets would have agglomerated from icy grains that remained pristine when entering the nebula. In the second case, comets would have agglomerated from icy grains condensed in the protosolar nebula and that would have been efficiently irradiated during their turbulent transport towards the upper layers of the disk. Both scenarios are found} consistent with the presence of molecular oxygen in comets.

\end{abstract} 


\section{Introduction}

The nature of the source of sulphur dimers (S$_2$) observed in comets is still unknown. The first detection of S$_2$ in a celestial body was in the UV spectra of Comet IRAS-Araki-Alcock (C/1983 H1) acquired with the International Ultraviolet Explorer (IUE) space observatory (A'Hearn et al. 1983). Emission bands of S$_2$ were subsequently identified in many comets observed with IUE in the eighties, including 1P/Halley (Krishna Swamy \& Wallis 1987). S$_2$ was also identified in comets Hyakutake (C/1996 B2), Lee (C/1999 H1), and Ikeya-Zhang (C/2002 C1) (Laffont et al. 1998; Kim et al. 2003; Boice \& Reyl\'e 2005). More recently, S$_2$ has been detected in comet 67P/Churyumov-Gerasimenko (hereafter 67P/C-G) by the ROSINA mass spectrometer aboard the Rosetta spacecraft at a distance of $\sim$3 AU from the Sun in October 2014 ($\sim$4--13 $\times$10$^{-6}$ with respect to water; Le Roy et al. 2015; {Calmonte et al. 2016}). All these observations suggest that S$_2$ is ubiquitous in comets.

Because the lifetime of S$_2$ is very short in comae ($\sim$a few hundreds seconds at most; Reyl\'e \& Boice 2003), two main scenarios have been invoked in the literature to account for its presence in comets. In the first scenario, S$_2$ is the product of reactions occurring in the coma. Ethylene was thus proposed to act as a catalyst allowing the formation of S$_2$ molecules in the inner coma (Saxena \& Misra 1995; Saxena et al. 2003). Also, the presence of atomic S (as the photodissociation product of CS$_2$) reacting with OCS was suggested to form S$_2$ in comae (A'Hearn et al. 2000). However, models depicting the chemistry occurring in cometary comae show that these two mechanisms do not account for the observed levels of S$_2$ (Rogers \& Charnley 2006).

In the second scenario, S$_2$ molecules are believed to be of parent nature and reside in cometary ices (A'Hearn et al. 1983; A'Hearn \& Feldman 1985; Grim \& Greenberg 1987; Feldman 1987; A'Hearn 1992). A'Hearn \& Feldman (1985) proposed that the UV photolysis of S-bearing species embedded in ISM ices could form sufficient amounts of S$_2$ that remains trapped in the icy matrix. Since then, a number of mechanisms based on UV or X-ray irradiation have been proposed, starting mainly from H$_2$S (the most abundant S-bearing volatile observed in comets; Irvine et al. 2000; Bockel\'ee-Morvan et al. 2004) and H$_2$S$_2$, and involving radicals like HS and HS$_2$ (Grim \& Greenberg 1987; Jim\'{e}nez-Escobar \& Mu\~{n}oz-Caro 2011; Jim\'{e}nez-Escobar et al. 2012). {It has also been proposed that S$_2$ could be formed from the radiolysis of S--bearing compounds in cometary ices (A'Hearn \& Feldman 1985; Calmonte et al. 2016) despite the fact that so far, there is no experimental proof showing that this mechanism is effective.

In the present study, we postulate that S$_2$ is formed from H$_2$S molecules embedded in icy grains by irradiation of UV, X-ray and cosmic ray fluxes
(CRF), whether icy grain precursors of comets formed in the protosolar nebula or the ISM.} Because radiolysis {generated by the impact of cosmic rays} simultaneously creates voids in ices within which the produced molecules can accumulate (Carlson et al. 2009; Mousis et al. 2016a), we investigate the stability of S$_2$ molecules in such cavities, assuming that the surrounding ice is made of H$_2$S or H$_2$O. We show that the stabilization energy of S$_2$ molecules in such voids is close to that of the H$_2$O ice binding energy, implying that they can only leave the icy matrix when this latter sublimates. We finally discuss the implications of our results for the origin of cometary grains, with a particular emphasis on those agglomerated by comet 67P/C-G.

\section{Irradiation of icy grains}

{Three irradiation mechanisms leading to the formation of S$_2$ are considered in this study. The first two mechanisms, namely UV and X-ray irradiation, have been proven to produce S$_2$ from H$_2$S and H$_2$S$_2$ (Grim \& Greenberg 1987; Jim\'{e}nez-Escobar \& Mu\~{n}oz-Caro 2011; Jim\'{e}nez-Escobar et al. 2012). Experiments have shown that S$_2$ can be produced and stabilized in icy grains over thicknesses of a few tenths of microns. Despite the lack of experimental data, radiolysis has also been considered as a potential candidate for S$_2$ formation from S--bearing compounds in cometary icy grains (A'Hearn \& Feldman 1985). This mechanism has recently been proposed to explain the detection of S$_2$ in 67P/C-G (Calmonte et al. 2016) and is often invoked to account for its presence in Europa's exosphere (Carlson et al. 1999; Cassidy et al. 2010). Cosmic rays reach deeper layers than photon irradiation and simultaneously creates voids in which some irradiation products such as O$_2$ or here S$_2$ can be sequestrated (Mousis et al. 2016a). Whatever the irradiation process considered, we assume that, once S$_2$ has been created and trapped in the microscopic icy grains, the latter agglomerated and formed the building blocks of comets.}

\section{Stability of S$_2$ molecules in an icy matrix}

The S$_2$ stabilization energy arises from the electronic interaction between the host support (H$_2$O ice or H$_2$S ice) and the S$_2$ foreign body.  The stabilization energy is evaluated as 

\begin{equation}
E_{stab} = (E_{ice} + E_{S_2}) - E, 
\end{equation}   
 
\noindent where $E_{S_2}$ is the energy of the isolated molecule, $E_{ice}$ the energy of the pristine solid host and $E$ is the  total energy of the [host + S$_2$] complex, with all entities optimized in isolation. 

All simulations are carried out by means of  the Vienna ab-initio simulation package (VASP) (Kresse \& Hafner 1993; 1994; Kresse \& {Furthm\"{u}ller 1996; Kresse \& Joubert 1999). The long range interactions in the solid and the hydrogen bonding being the critical parameters in the ices, we use the PBE generalized gradient approximation (GGA) functional (Perdew et al 1996), in the (PBE+D2) version corrected by Grimme et al. (2010) that has been specifically designed to deal with the present type of problem. This theoretical tool has proved to be well adapted to model bulk and surface ice structures interacting with volatile species (Lattelais et al.  2011, 2015; Ellinger et al. 2015; Mousis et al. 2016a). More details on the computational background can be found in the aforementioned publications.

Since S$_2$ is created well inside the icy grain mantles, the initial description of the irradiated ice is taken as that of the internal structures of ice clusters obtained from Monte-Carlo simulations of ice aggregates constituted of hundreds of water molecules.} The important point in the simulations by Buch et al (2004) is that the core of the aggregates consists in crystalline domains of apolar hexagonal ice {\it Ih}. However, in the present context, the irradiation creates significant defects inside  the ice, namely, voids and irradiation tracks that, at least locally, modify the crystalline arrangement.  

\subsection{S$_2$ embedded in H$_2$O  ice }

Because H$_2$O is the dominant volatile in comets (Bockel\'ee-Morvan et al. 2004), most of the cavities created by CRF irradiation are expected to be surrounded by H$_2$O molecules. Table 1 shows the stabilization energy of S$_2$ as a function of the size of these cavities. How the S$_2$ stabilization evolves as a function of their size is summarized below. 

\begin{enumerate}
\item Starting with no H$_2$O removed, we find no stabilization for the inclusion of S$_2$ in the ice lattice. It is in fact an endothermic process, as it is for O$_2$ inclusion (Mousis et al. 2016a).

\item With one H$_2$O removed, we have an inclusion structure whose stabilization is negative, meaning that S$_2$ cannot stay in such a small cavity. 

\item With somewhat larger cavities obtained by removing 2 to 4 adjacent H$_2$O molecules from the ice lattice, we obtain increasing stabilization energies from 0.3 to 0.5 eV.

\item With larger cavities that form along the irradiation track, the stabilization energies are found to be at least of the order of 0.5 eV.

\end{enumerate}

In short, as soon as the space available is sufficient, the energy stabilizes around 0.5 eV. This stabilization energy is i) higher (more stabilizing) than what is found in the case of O$_2$ (0.2--0.4 eV; Mousis et al. 2016a) and ii) larger than that of a water dimer ($\sim$0.25 eV). Hence, the presence of S$_2$  should not perturb the ice structure until it is ejected into the coma via sublimation with the surrounding H$_2$O molecules. The results of our computations are consistent with the laboratory experiments of Grim \& Greenberg (1987) who showed that S$_2$ remains trapped in icy grains until they are heated up to $\sim$160 K, a temperature at which water ice sublimates at PSN conditions.

\begin{table}[h]
\begin{center}
\caption[]{Computed stabilization energies (eV) of S$_2$ interacting 	with H$_2$O ice or H$_2$S ice}
\begin{tabular}{lcc}     
\hline
\hline                            			
Environment              			& 	H$_2$O ice    	&  	H$_2$S ice     \\
\hline                                        
Adsorption  	     				&   	0.28 			&				\\
Inclusion (n=1)$^\ast$ 			&   	-0.12 		&   	0.30			\\
Inclusion (n=2)    				&   	0.28    		&   	0.45			\\
Inclusion (n=4)     				&   	0.50 	         	&   	0.40			\\
Inclusion (fine track)				& 	0.51           	&   	0.41			\\ 
Inclusion (large track)			& 	0.53 	          	&   	0.50  		\\
\hline                  
\end{tabular}
\end{center}
$^\ast$n = number of H$_2$O or H$_2$S molecules destroyed to create the void in which  S$_2$ is trapped.
\label{tab1}
\end{table}


\subsection{S$_2$ embedded in H$_2$S  ice}

H$_2$S behaves similarly to H$_2$O because of its ability to establish hydrogen bonds. This implies that small domains of H$_2$S could have formed in the bulk of the ice and served as local sources for the formation of  S$_2$. The stabilization of these aggregates is addressed by numerical simulations in which H$_2$S entities are progressively introduced by replacing an equal number of H$_2$O molecules in the water-ice lattice. Table 2 shows the stabilization energies with values around 0.5 and 0.75 eV for neighboring and far away H$_2$S, respectively. Consequently, substituting several neighboring H$_2$O by H$_2$S is a possibility to be considered if the H$_2$S is abundant enough, thus creating small islands of H$_2$S within the water ice.

\begin{table}[h]
\begin{center}  
\caption{Computed stabilization energies (eV) of H$_2$S interacting with H$_2$O ice}         
\begin{tabular}{lc}     
\hline
\hline       
                      						& SH$_2$   	\\ 
Environment              				& E$_{stab}$     \\
\hline                                        	
Adsorption  	     					&   	0.61   	\\
Substitution (n=1)$^\ast$   			&   	0.77		\\
Substitution (n=2 far away)     			&   	0.73    	\\
Substitution (n=2 close)     			&   	0.56    	\\
Substitution (n=3 close)     			&   	0.50   	\\
Substitution (irradiation track)    		&   	0.51  	\\
\hline                  
\end{tabular}
\end{center}
$^\ast$n = number of H$_2$O molecules replaced by H$_2$S. 
\label{tab2}
\end{table}

If small clumps of H$_2$S ices in the bulk of water ice are a plausible hypothesis, as suggested by the aforementioned numbers, then the proper conditions are realized for the in situ formation of S$_2$ by deep irradiation. The case in which H$_2$S molecules replace H$_2$O along the irradiation track is a less favorable situation but it could be at the origin of the S$_n$  oligomers observed in some laboratory experiments (Meyer et al. 1972;  Jim\'{e}nez-Escobar et al. 2012). We evaluate the stabilization of S$_2$ in H$_2$S clumps, assuming that they behave as pure condensates. The results, presented in Table 1 and summarized below, are quite close to those derived for water ice.

\begin{enumerate}
\item With one H$_2$S removed, we have a substitution structure whose stabilization is on the order of 0.30 eV. 

\item With larger cavities obtained by removing 2 to 4 adjacent H$_2$S molecules, we obtain increasing stabilization energies between 0.40 and 0.45 eV.

\item With even larger cavities, extended in the direction of the irradiation, the stabilization energies are found to be similar to the preceding ones, between 0.40 and 0.50 eV.

\end{enumerate}

Again we find that the presence of S$_2$  should not perturb the ice structure, even when trapped in H$_2$S clumps, until the latter sublimate, due to increasing local temperature.

\section{Implications for cometary ices}

It has been recently shown that the radiolysis of icy grains in low-density environments such as the presolar cloud may induce the production of amounts of molecular oxygen high enough to be consistent with the quantities observed in 67P/C-G (Mousis et al. 2016a). Higher density environments such as the PSN midplane were excluded because the timescales needed to produce enough O$_2$ in cometary grains exceeded by far their lifetimes in the disk. Also, the efficiency of ionization by cosmic rays in the PSN midplane is now questioned because of the deflection of galactic CRF by the stellar winds produced by young stars (Cleeves et al. 2013, 2014).

On the other hand, because the lifetime of S$_2$ is very short in the gas phase ($\sim$a few hundred seconds at most; Reyl\'e \& Boice 2003), its formation conditions are even more restrictive than those required for O$_2$. Assuming that S$_2$ indeed formed from H$_2$S {or any other S-bearing molecule} via {UV, X-ray or CRF irradiation}, this implies that this molecule never left the icy matrix in the time interval between its formation and trapping. In other words, S$_2$ never condensed from the PSN before being trapped in cometary grains. This stringent constrain requires S$_2$ to form within icy grains irradiated by CRF in low density environments such as ISM, where the local temperature is extremely cold. In this picture, comets, including 67P/C-G, would have agglomerated in the PSN from icy grains originating from ISM, whose compositions and structures remained pristine when entering the nebula.

Alternatively, because the CRF irradiation should be poorly attenuated in the upper layers of the PSN, these regions also constitute an adequate low-density environment allowing the formation of S$_2$ in cometary grains. Turbulence plays an important role in the motion of small dust grains that are well coupled to the gas. Micron-sized grains initially settled in the midplane are  entrained by turbulent eddies and diffuse radially and vertically with an effective viscosity roughly equal to that of the gas for such small particles (see Ciesla 2010, 2011 for details). Consequently, solid particles follow a gaussian distribution in the vertical direction. The scale height of dust (corresponding to the standard deviation of the distribution) is a fraction of the gas scale height, this fraction being larger and possibly equal to the gas scale height in the cases of small grains and higher degrees of turbulence (Dubrulle et al. 1995; Youdin \& Lithwick 2007).

The vertical transport of solids exposes them to very different disk environments. Dust grains are stochastically transported to high altitude and low-density regions above the disk midplane. Ciesla (2010) developed a numerical simulation to integrate the motion of individual particles and showed that micron-sized grains spent $\sim$32\% of their lifetime at altitudes above the scale height of the disk, including $\sim$5\% at heights above four times its scale height, regardless the distance from the Sun. In such low-density environments, photochemistry plays a primordial role, as demonstrated by Ciesla \& Sandford (2012), because UV photons are weakly attenuated at those heights. This also holds for the CRF irradiation of grains that should be substantially enhanced compared to the dose received by particles residing in the midplane. Under those circumstances, the production of S$_2$ should be favored in icy grains over several cycles of vertical transport towards the surface of the disk. This scenario should also favor the formation of O$_2$ from irradiation of H$_2$O ice (see Mousis et al. 2016a for details).

\begin{figure}[h]
\begin{center}
\resizebox{\hsize}{!}{\includegraphics[angle=0]{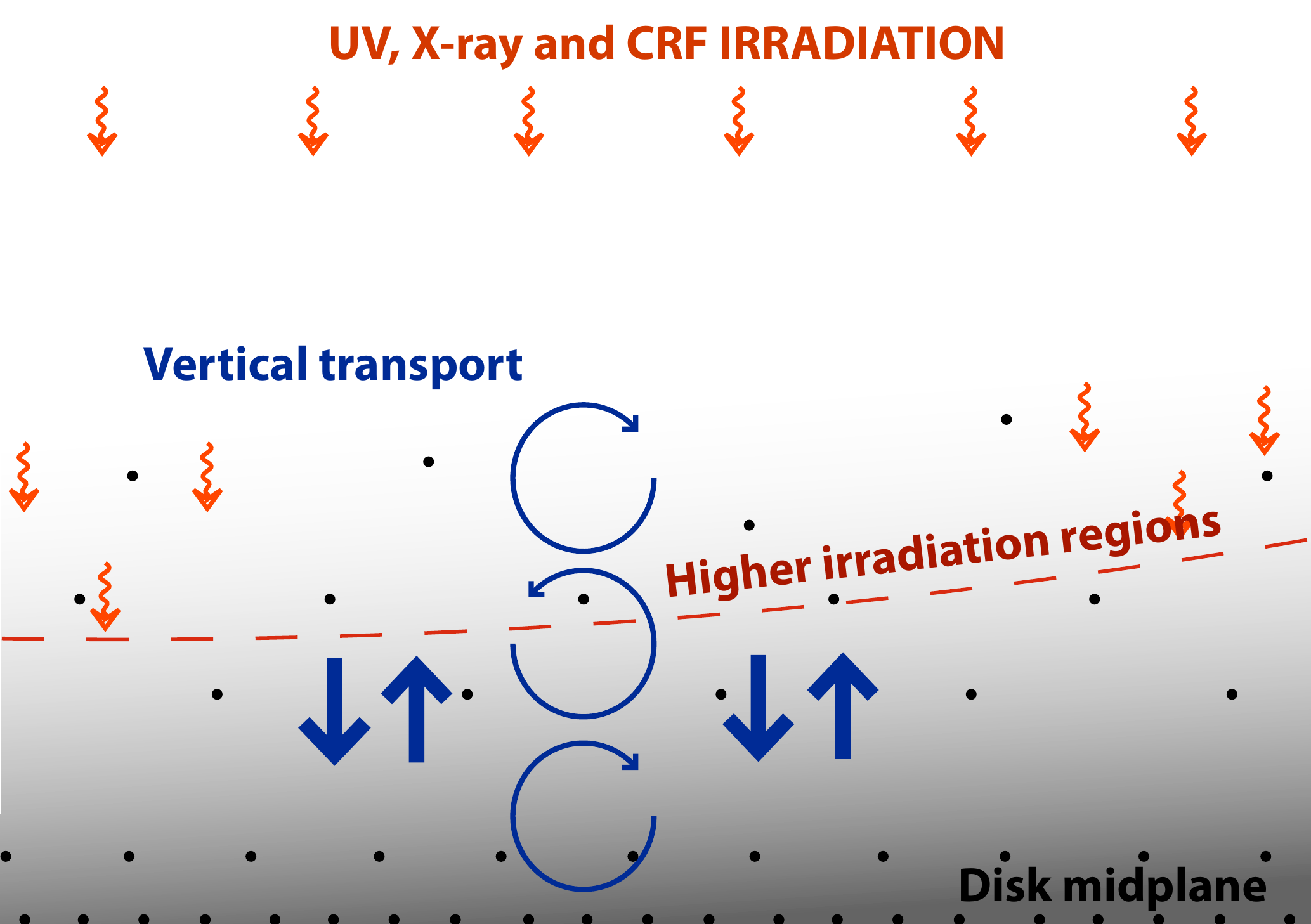}}
\caption{Illustration of the vertical transport of small icy grains towards disk regions where they are efficiently {irradiated}. Dust is concentrated in the midplane of the disk due to gravitational settling and gas drag. However, turbulent eddies lift the icy grains toward the upper regions and also drag them down as the direction of the velocity is random and coherent during a timecale comparable to the local keplerian period. Small dust grains finally spend a non negligible fraction of their lifetime in the disk's upper regions, where the {irradiation attenuation is low}.}
\label{holes}
\end{center}
\end{figure}

\section{Discussion and conclusions}

{It is reasonable to assume that the multiple forms of irradiation hitting the microscopic icy grains in low density environments such as ISM or the upper layers of protoplanetary disks can lead both to the formation of S$_2$ molecules and the development of cavities in these grains, in which the molecule remains sequestered. The same scenario has been proposed for O$_2$ formation and stabilization in cometary icy grains (Mousis et al. 2016a). In the case of S$_2$ formation,  the possibility of forming the dimer via the radiolysis of S-bearing ices remains an open question. Future experimental work is needed to check the viability of this mechanism.}

The possible formation of S$_2$ in icy grains via their {irradiation} in ISM, together with the short lifetime of this molecule in the gas phase, leads to the plausible possibility that comets agglomerated from pristine amorphous grains that never vaporized when entering the PSN, as already envisaged for the origin of 67P/C-G's material (Rubin et al. 2015a; Mousis et al. 2016a). On the other hand, the formation of S$_2$ in icy grains that migrated towards the upper layers of the disk is compatible with their condensation in the PSN midplane. This mechanism leaves open the possibility that these grains are made of crystalline ices and clathrates, as proposed by Mousis et al. (2016b) and Luspay-Kuti et al. (2016) to account for several pre-perihelion compositional measurements made by the Rosetta spacecraft in 67P/C-G. The same process could explain the presence of O$_2$ measured in situ in comets 67P/C-G and 1P/Halley (Bieler et al. 2015; Rubin et al. 2015b). {Interestingly, whatever the ice structure considered for the icy grains, the voids allowing the stabilization of S$_2$ can be considered as analogs of clathrates in terms of cage sizes and intermolecular interactions.}

The fact that one H$_2$S replacing one H$_2$O has little influence on the stability of the solid lattice is a favorable situation for the formation of a mixed ice. It is  plausible that some segregation occurs with the formation of H$_2$S islands in the bulk of crystalline or amorphous water ice. Then, the proper conditions would be realized for the in situ formation of S$_2$, especially if  we remember that the formation of one S$_2$ requires at least the destruction of two imprisoned sulphur species. The plausible formation of H$_2$S clumps is a strong argument in favor of a non uniform distribution of S$_2$ within cometary ices. Note that in the case of irradiation of crystalline grains condensed in the PSN and transported towards the upper layers of the disk, the formed S$_2$ may be entrapped in clathrates (Grim \& Greenberg 1987), also forming a solid phase distinct from water ice in cometary grains.

The immediate consequence of the presence of distinct S$_2$-bearing solid phases is the difficulty to predict the S$_2$ correlation with {H$_2$O or H$_2$S} in 67P/C-G from Rosetta measurements. The S$_2$/H$_2$O abundance ratio is directly linked to the region of the comet whose desorption is observed. Contrary to O$_2$ whose apparent good correlation with H$_2$O is explained by its trapping in water ice (Bieler et al. 2015; Mousis et al. 2016a), no global trend should be drawn between the variation of S$_2$ and H$_2$O abundances if S$_2$ is distributed within both the S-bearing and H$_2$O ices. Indeed, S$_2$ may be released simultaneously from the H$_2$O layer present close to the surface and from H$_2$S clusters localized deeper in the subsurface. {Our results are supported by the ROSINA data collected between May 2015 (equinox) and August 2015 (perihelion), showing that there is no clear correlation of S$_2$ with H$_2$O or H$_2$S in 67P/C-G (Calmonte et al. 2016). These observations allow us to exclude the trapping of S$_2$ in a dominant ice reservoir. If S$_2$ was mainly trapped in H$_2$S--bearing ice, then the outgassing rates of S$_2$ and H$_2$S should have been well correlated during the period sampled by the ROSINA instrument. The same statement applies if S$_2$ had been essentially trapped in water ice.}

\acknowledgements
O.M. acknowledges support from CNES. This work has been partly carried out thanks to the support of the A*MIDEX project (n\textsuperscript{o} ANR-11-IDEX-0001-02) funded by the ``Investissements d'Avenir'' French Government program, managed by the French National Research Agency (ANR). This work also benefited from the support of CNRS-INSU national program for  planetology (PNP). J.I.L appreciates support from NASA through the JWST project. A.L.-K acknowledges support from NASA  JPL (subcontract no. 1496541).

\end{document}